\documentclass{IEEE_lsens}
\usepackage{textcomp}
\usepackage{graphicx}

\usepackage[noadjust]{cite}
\ifCLASSINFOpdf
\else
\fi
\usepackage{amsmath,amssymb,amsfonts}
\usepackage{algorithm}
\usepackage{algpseudocode}
\usepackage{graphicx}
\usepackage{textcomp}
\usepackage{newunicodechar}
\newunicodechar{​}{} % U+200B Zero Width Space

\usepackage{hyperref}
\usepackage{xcolor}
\DeclareMathOperator{\atantwo}{atan2}

\algrenewcommand\algorithmicrequire{\textbf{Input:}}
\algrenewcommand\algorithmicensure{\textbf{Output:}}

\usepackage[T1]{fontenc} % optional enhanced font encoding
\usepackage{amsmath}
\usepackage[cmintegrals]{newtxmath}
\usepackage{bm}
\usepackage{array}
\usepackage{url}
\ifCLASSINFOpdf
\else
\fi
\providecommand{\hypersetup}[1]{\relax}

\hypersetup{pdftitle={Bare Demo of IEEE\_lsens.cls for IEEE Sensors Letters},%<!CHANGE!
	pdfsubject={Typesetting},%<!CHANGE!
	pdfauthor={Talha Nadeem, Muhammad Tahir},%<!CHANGE!
	pdfkeywords={Class, IEEE, IEEE\_lsens, IEEE Sensors Letters, LaTeX, Typesetting, TeX}}%<^!CHANGE!}
\begin{document}

% article subject line
\IEEELSENSarticlesubject{Sensor signals processing}

\title{NIR-EKF: Normalized Innovation Ratio-Based EKF for Robust State Estimation}

\author{%
	\IEEEauthorblockN{Talha~Nadeem\IEEEauthorrefmark{1},
		Khurram~Ali\IEEEauthorrefmark{2},
		and~Muhammad~Tahir\IEEEauthorrefmark{1}}%
	\IEEEauthorblockA{%
		\IEEEauthorrefmark{1}Department of Electrical Engineering,
		Lahore University of Management Sciences, Lahore 54792, Pakistan\\
		\IEEEauthorrefmark{2}Department of Electrical, Computer Engineering,
		COMSATS University Islamabad, Lahore 54000, Pakistan\\
		\IEEEauthorieeemembermark{1}Senior Member, IEEE}%
}
%
% note the % following lines that end in } -
% these prevent an unwanted space from occurring between the objects.
% i.e., if you had this:
% For what it is worth, this is a minor point as most people would not even
% notice if the said evil space somehow managed to creep in.

% Manuscript received line
%\IEEELSENSmanuscriptreceived{Manuscript received June 7, 2017;
%revised June 21, 2017; accepted July 6, 2017.
%Date of publication July 12, 2017; date of current version July 12, 2017.}

% for Sensors Letters papers, we must declare the abstract and index terms
\IEEEtitleabstractindextext{%
\begin{abstract}
	Sensors deployed in real-world conditions often produce measurements corrupted by
	outliers due to model uncertainties, changes in the surrounding environment, and/or
	data loss. As a result, managing these outliers becomes crucial for state estimation
	to avoid inaccurate estimations and a reduction in the reliability of results. To
	address this issue, we introduce a novel form of extended Kalman filter (EKF) based
	on the maximum a posteriori (MAP) principle for scenarios where outliers
	simultaneously occur in multiple dimensions. For detecting outliers during the
	filtering process, we introduce a novel variant of the normalized innovation ratio
	(NIR) test and embed it within the EKF framework. Our approach enhances the
	estimation accuracy and computational efficiency of state estimation process even
	when data from several sensors simultaneously contain outliers.
\end{abstract}

\begin{IEEEkeywords}
	Sensor signal processing, extended Kalman filter (EKF), maximum a posteriori
	(MAP), normalized innovation ratio (NIR), outliers, reduced sequential maximum
	a posteriori-based (RSMAP), sensors signal processing, state estimation.
\end{IEEEkeywords}}

\maketitle

\section{Introduction}

State estimation is a critical problem in engineering and statistics,
fundamental to the operation and control of systems in diverse domains,
such as power systems \cite{paul2018centralized}, robotics
\cite{barfoot2017state}, and target tracking
\cite{gu2015gnss, ivanovstate2015}. At the heart of this problem lies the
challenge of accurately predicting system states, a task for which the
Kalman filter (KF) and its nonlinear version, the extended Kalman filter
(EKF) \cite{faragher2012understanding, liu2016extended}, have been widely
adopted. These filters operate under the assumption that noise and system
dynamics can be modeled as Gaussian processes, making them exceptionally
effective under linear conditions or when nonlinearities can be adequately
linearized. However, real-world use of the KF family faces challenges from
outliers, i.e., measurements that are significantly different from expected
distribution due to sensor issues, environmental disruptions, malicious
activities, or data errors \cite{fang2018robustifying, li2022novel,
	li2017detection}. These anomalies can drastically affect state estimation
algorithms, resulting in inaccurate predictions.

Outliers may distort data from a single sensor or simultaneously across
multiple sensors. Therefore, the strategies for outlier mitigation fall
into two categories as well. Some of the initial works, which aim toward
managing outliers occurring in a single dimension, are based on robust
M-Estimation methods. These methods were developed to mitigate outlier
effects by adjusting their influence during the estimation process
\cite{durovic1999robust}, \cite{ting2007kalman}. Concurrently, $l_1$ norm
filters approach outlier handling through optimization strategies
\cite{graham2015robust}, although their performance can be limited in
dynamic environments. Another family of methods, $H_{\infty}$ filtering,
aims at ensuring robustness against extreme scenarios \cite{li2016h},
yet it may introduce a conservative bias. The main limitation of these
methods is their mitigation of outliers in only one dimension.

Progressing from these foundational approaches, recent research works have
explored non-Gaussian models to represent outliers. These methods mainly
deal with outliers occurring in multiple dimensions but had other major
drawbacks. As an example, some of the works use the Monte Carlo method
\cite{kitagawa1996monte} and Student-t distributions
\cite{li2018robust}, which require precise parameter tuning for optimal
performance despite offering enhanced flexibility. Variational-Bayes (VB)
techniques address the identification of outliers in multiple sensors
through a probabilistic framework, although with challenges, such as slow
convergence and high computational needs, due to iterative optimization per
step for VB \cite{wang2018robust, li2020robust}. The evolution of outlier
mitigation, incorporating advanced filtering, such as the enhanced EKF
\cite{mu2015novel} and robust variational Bayesian filter
\cite{huang2023novel}, has improved robustness, however, these methods,
while effective in multiple dimension outlier rejection, often struggle
with the balance between computational efficiency and the risk of
discarding valid data. Techniques, such as regression-based exclusion
\cite{de2017output} and the cubature Kalman filter \cite{zhang2022minimum},
also face similar tradeoffs, emphasizing the ongoing challenge in outlier
detection and mitigation in both single and multiple dimensions of
measurement vector.

In this letter, we propose a novel robust filtering method where we use an
explicit model for outlier occurrence in measurement dimensions followed
by maximum a posteriori (MAP)-based inference. In order to handle all
measurement dimensions, the estimator is modified to evaluate only two
probabilities per dimension, which reduces the computational cost
significantly. Further, for detection of outliers, we derive an equivalence
between the normalized innovation ratio (NIR) test
\cite{graham2015robust, qi2018comparing} and our proposed MAP-based
estimator, providing us a very simple rule to detect the presence of
outliers. We call our method the NIR-EKF because it uses an NIR-based
threshold for outlier detection within the proposed MAP-based EKF. We
further evaluate our algorithm against robust filters for accuracy and
computational efficiency. Our approach has the following key advantages
over prior works.

\begin{enumerate}
	\item It is fast and has extremely low computational cost.
	\item It handles the outliers in multiple dimensions.
	\item It is robust to outliers.
	\item It is easy to implement and straightforward to apply.
\end{enumerate}
\section{Proposed Method - NIR-EKF}
%\subsection{Organization}
%We present a formal structure for the problem of outliers in measurement sensors. Then we recount the result of \cite{chughtai2020dynamic} in an overview followed by a description of our proposed estimators.
%\subsection{Problem Formulation}
Consider the following dynamical system:
\begin{align}
	\mathbf{x}_k&=\mathbf{f}(\mathbf{x}_{k-1})+\mathbf{q}_{k-1} \label{eqn1}\\
	\mathbf{y}_k&=\mathbf{h}(\mathbf{x}_{k})+\mathbf{r}_k+\mathbf{v}_k \label{eqn2}
\end{align}
where $\mathbf{x}_k\in\mathbb{R}^n$ is the state vector at the time instant $k$, and $\mathbf{y}_k\in\mathbb{R}^m$ is the measurement vector. The functions $\mathbf{f(.)}:\mathbb{R}^n\rightarrow\mathbb{R}^n$ and $\mathbf{h(.)}:\mathbb{R}^n\rightarrow\mathbb{R}^m$ define the process dynamics and observation equations, respectively, and the vectors $\mathbf{q}_{k-1}$ and $\mathbf{r}_k$ are the state and process noise vectors, respectively, modeled by multivariate normal distributions with covariances $\mathbf{Q}_k\in\mathbb{R}^{n\times n}$ and $\mathbf{R}_k\in\mathbb{R}^{m\times m}$. In addition to the process noise in the measurement vector, we model measurements corrupted with outliers by adding the vector $\mathbf{v}_k$.

\subsection{MAP-EKF}

If we treat $\mathbf{v}_k$ as a multimodal Gaussian distribution as
\begin{align}
	&p(\mathbf{v}_k) = { \sum_{i=0}^{m} p(\mathbf{v}_k | I^i)  p(I^i)}  \label{eqn3}\\
	&p(\mathbf{v}_k| I^0) = { \delta (\mathbf{v}_k)  } \label{eqn4}\\
	&p(\mathbf{v}_k| I^i) = { \mathcal{N} (\mathbf{v}_k;\mathbf{0}_k^i, \mathbf{\Sigma}^i)  } \label{eqn5}
\end{align}
Expressing $\mathbf{v}_k$ as in \eqref{eqn3}--\eqref{eqn5} captures different measurement scenarios, such as the presence or absence of outliers. The event $I^0$ represents no outliers in any dimension as modeled by \eqref{eqn4} implying no additional noise from outliers in the measurement, while $I^i$ modeled by \eqref{eqn5} highlights an outlier in the $i^{th}$ dimension. The vector $\mathbf{0}_k^{i}$ is a mean vector with all zero elements except the $i^{th}$ element, which is set to $0^i$. The covariance matrix $\mathbf{\Sigma}^i$ features a nonzero entry only on the $i^{th}$ diagonal, increasing the variance for the $i^{th}$ dimension's measurement in the filtering process. Moreover, the symbol
\[
\mathcal{N}(\mathbf{x};\boldsymbol{\mu},\boldsymbol{\Sigma})
=
\frac{1}{\sqrt{|\boldsymbol{\Sigma}|(2\pi)^n}}
\exp\left(
-\frac{1}{2}
(\mathbf{x}-\boldsymbol{\mu})^\top
\boldsymbol{\Sigma}^{-1}
(\mathbf{x}-\boldsymbol{\mu})
\right)
\]
represents the standard multinomial Gaussian. Using the law of total probability, we obtain
\begin{equation}
	p(\mathbf{v}_k) =
	\pi^0 \delta(\mathbf{v}_k)
	+
	\sum_{i=1}^{m}
	\pi^i
	\mathcal{N}(\mathbf{v}_k;\mathbf{0}_k^i,\mathbf{\Sigma}^i)
	\label{eqn6}
\end{equation}
where $\pi^0$ denotes the prior probability corresponding to the event $I^0$, while $\pi^i$ is the prior probability for the event $I^i$. Under this model, we can obtain the state posterior, accounting for all $I^i$ events where $i\in\{0,1,\ldots,m\}$, as follows \cite{chughtai2020dynamic}
\begin{align}
	p(\mathbf{x}_k={\boldsymbol{\mu}_k^i}|\mathbf{y}_{1:k})
	&\propto
	\sum_{i=0}^{m}
	{\mathcal{N}(\mathbf{x}_k;\boldsymbol{\mu}_k^i,\mathbf{P}_k^i)
		\mathcal{N}(\mathbf{y}_k;\mathbf{\bar{y}}^i,\mathbf{S}_k^i)}
	\pi^i
	\label{eqn8n}
\end{align}
where
$\boldsymbol{\mu}_k^- = \mathbf{f}(\boldsymbol{\mu}_{k-1}^+)$,
$\boldsymbol{\mu}_k^i=\boldsymbol{\mu}_k^-+\mathbf{K}_k^i(\mathbf{y}_k-\mathbf{\bar{y}}_k^-)$,
$\mathbf{\bar{y}}_k^-=\mathbf{h}(\boldsymbol{\mu}_k^{-})$,
$\mathbf{S}_k^{i}=\mathbf{H}(\boldsymbol{\mu}_k^-)\mathbf{P}_k^-
\mathbf{H}^T(\boldsymbol{\mu}_k^-)+\mathbf{R}_k+\mathbf{\Sigma}^i$,
$\mathbf{C}_k=\mathbf{P}_k^{-}\mathbf{H}^T(\boldsymbol{\mu}_k^-)$,
$\mathbf{K}_k^i=\mathbf{C}_k\mathbf{S}_k^{-1}$
and
$\mathbf{P}_k^i=
\mathbf{P}_k^{-}+
\mathbf{K}_k^i\mathbf{S}_k^i\mathbf{K}_k^{i^T}$.
Moreover, the event $I^{0}$ denotes no outlier in any of the measurements; therefore, $\mathbf{\Sigma^0}=[0]_{m\times m}$. This zero matrix indicates no additional variability from outliers, aligning with the definition of $I^0$. This process results in a set of candidate estimates for the state vector
$\{\boldsymbol{\mu}_{k}^{0},\boldsymbol{\mu}_{k}^{1},
\boldsymbol{\mu}_{k}^{2},\ldots,\boldsymbol{\mu}_{k}^{m}\}$.

For time step $k$, the Maximum A Posteriori (MAP) estimator selects the estimate with the highest posterior probability from index $j$ as
$j=\arg\max_i p(\mathbf{x}_k=\boldsymbol{\mu}_k^i|\mathbf{y}_{1:k})$,
where $\boldsymbol{\mu}_k^+$ is set to $\boldsymbol{\mu}_k^j$,
$\mathbf{P}_k^+$ is updated to $\mathbf{P}_k^j$, and the estimated state
$\mathbf{\hat{x}}_k$ becomes $\boldsymbol{\mu}_k^+$.
\subsection{Exhaustive MAP-EKF}

We need to remodel the decomposition of the distribution of
$\mathbf{v}_k$ for the independent occurrence of outliers in the
measurements. Let $\Gamma$ denote all the possible subsets of sensors
of an agent. All the elements of $\Gamma$ are mutually exclusive, but
$\Gamma$ is an exhaustive set of all the $2^m$ possible distinct subsets
of sensors that can have outliers. Hence, we can rewrite
\eqref{eqn3}--\eqref{eqn5} as
\begin{align}
	p(\mathbf{v}_k) &= \sum_{\gamma \in \Gamma}
	p(\mathbf{v}_k | I^\gamma)p(I^\gamma)
	\label{8}\\
	p(\mathbf{v}_k | I^\gamma) &=
	\mathcal{N}(\mathbf{v}_k; \mathbf{0}^\gamma,
	\mathbf{\Sigma}^{\gamma})
	\label{9}\\
	p(\mathbf{v}_k) &= \prod_{\gamma \in \Gamma}
	\pi^\gamma \mathcal{N}(\mathbf{0}^\gamma,
	\mathbf{\Sigma}^{\gamma})
	\label{10}
\end{align}

In \eqref{8}--\eqref{10}, $I^\gamma$ signifies an outlier in sensor
subset $\gamma$, where $\mathbf{0}^\gamma$ and
$\mathbf{\Sigma}^{\gamma}$'s diagonal hold nonzero values only at
indices for sensors within $\gamma$. For inference, an eventual
improvement to the MAP-EKF is an exhaustive MAP-EKF (EMAP-EKF),
which checks for all possible combinations of the outliers in
measurements, i.e., $2^m$. It considers $2^m$ scenarios for $m$
measurements to produce
$\{\boldsymbol{\mu}_{k}^{0}, \boldsymbol{\mu}_{k}^{1},
\boldsymbol{\mu}_{k}^{2}, \ldots,
\boldsymbol{\mu}_{k}^{2^m}\}$ estimates. Similar to the MAP-EKF, the
EMAP-EKF uses a MAP estimator to select the most appropriate estimate.

\subsection{Sequential MAP-EKF}

The EMAP-EKF's major downside is its $\mathcal{O}(2^m)$ computational
cost. To reduce this cost, we introduce an alternative estimator that
handles $m$ measurements, requiring two assessments each (as outlier
or nonoutlier), leading to a $\mathcal{O}(m)$ time complexity. This
estimator sequentially processes the measurement vector $\mathbf{y}_k$,
where each element $y_k^i$ informs the prediction for the next, using
two filters per dimension to accommodate possible outliers. This
approach yields a pair of estimates
$\{\boldsymbol{\mu}_{0}^{i}, \boldsymbol{\mu}_{C}^{i}\}$ per dimension,
from which the MAP estimator selects the most likely. This modifies
the process previously described by \eqref{eqn8n}.
\begin{align}
	p(\boldsymbol{\mu}_j^{i}|y_{k}^{i}) \propto
	&\mathcal{N}(\boldsymbol{\mu}_j^i;
	\boldsymbol{\mu}_0^i,\mathbf{P}_0^i)
	\mathcal{N}(y_k^{i};\bar{y}_k^i,S_k^0)\pi^0
	\nonumber\\
	&+\mathcal{N}(\boldsymbol{\mu}_j^i;
	\boldsymbol{\mu}_C^i,\mathbf{P}_C^i)
	\mathcal{N}(y_k^{i};\bar{y}_k^i,S_k^C)\pi^C
	\label{11}
\end{align}

The selection of the estimate is determined by the value of $c$ as
determined by
$c = \underset{j \in \{0,C\}}{\arg\max}\:
p(\boldsymbol{\mu}_j^{i}|y_{k}^{i})$, after which
$\boldsymbol{\mu}^{i}$ is set to $\boldsymbol{\mu}_c^i$ and
$\mathbf{P}^{i}$ is updated to $\mathbf{P}_c^{i}$. Here,
$\pi^{0}$ and $\pi^{C}$ denote the prior probabilities for
$I_i^{0}$ and $I_i^{C}$, respectively. The term
$\bar{y}_k^i = h_i(\boldsymbol{\mu}^{i-1})$, where
$h_i(\boldsymbol{\mu}^{i-1})$ is the $i^{th}$ component of
$h(\boldsymbol{\mu}^{i-1})$. To account for increased variance when
an outlier is present, we use
$S_k^{C} = S_k^{0} + \Sigma^{C}$.
The updated estimate is given by
$\mathbf{x}_k^{+} = \boldsymbol{\mu}^{m}$.
As this method processes the measurement vector in a sequential
manner, we call it the sequential MAP-EKF (SMAP-EKF).
In this filter, the MAP estimator makes a choice between the
following two probabilities:

\begin{align}
	p(\boldsymbol{\mu}_0^{i}|y_{k}^{i}) \propto
	&\mathcal{N}(\boldsymbol{\mu}_0^i;
	\boldsymbol{\mu}_0^i,\mathbf{P}_0^i)
	\mathcal{N}(y_k^{i};\bar{y}_k^i,S_k^0)\pi^0
	\nonumber\\
	&+\mathcal{N}(\boldsymbol{\mu}_0^i;
	\boldsymbol{\mu}_C^i,\mathbf{P}_C^i)
	\mathcal{N}(y_k^{i};\bar{y}_k^i,S_k^C)\pi^C
	\label{12}
\end{align}

\begin{align}
	p(\boldsymbol{\mu}_C^{i}|y_{k}^{i}) \propto
	&\mathcal{N}(\boldsymbol{\mu}_C^i;
	\boldsymbol{\mu}_0^i,\mathbf{P}_0^i)
	\mathcal{N}(y_k^{i};\bar{y}_k^i,S_k^0)\pi^0
	\nonumber\\
	&+\mathcal{N}(\boldsymbol{\mu}_C^i;
	\boldsymbol{\mu}_C^i,\mathbf{P}_C^i)
	\mathcal{N}(y_k^{i};\bar{y}_k^i,S_k^C)\pi^C.
	\label{13}
\end{align}

\subsection{Reduced SMAP-EKF}

Assuming both $\pi^{0}$ and $\pi^{C}$ in \eqref{12} and \eqref{13}
equal to 0.5, the key determinant between two probabilities lies in
comparing $\mathcal{N}(y_k^{i};\bar{y}_k^i,S_k^0)$ with
$\mathcal{N}(y_k^{i};\bar{y}_k^i,S_k^C)$. If
$\mathcal{N}(y_k^{i};\bar{y}_k^i,S_k^0) >
\mathcal{N}(y_k^{i};\bar{y}_k^i,S_k^C)$, indicating that $y_k^{i}$
closely matches nominal characteristics, the MAP estimator selects
$\boldsymbol{\mu}_0^{i}$, otherwise $\boldsymbol{\mu}_C^{i}$.
This method, we term as reduced SMAP-EKF (RSMAP-EKF), performs two
evaluations per measurement ($m$) and has a computational cost of
$\mathcal{O}(m)$. The RSMAP-EKF is faster than the SMAP-EKF due to
its simplified probability function and assumptions.
\subsection{NIR-EKF}

Now, we further built on RSMAP-EKF to determine a way of detecting
the outlier implicitly. For this purpose, we establish an equivalence
relationship between the MAP estimator in the RSMAP-EKF and the NIR
test \cite{qi2018comparing} for the sensor measurements. The NIR test
detects outliers by normalizing the difference between actual and
predicted measurements. Our proposed method flags and discards
measurements exceeding a set threshold, ensuring robustness. For
instance, a faulty factory temperature sensor's readings are compared
to predicted values; significant deviations are flagged and ignored.
We start with the assertion that in the scenario where no outlier is
present, the following condition must be met:

\begin{align}
	\mathcal{N}(y_k^{i};\bar{y}_k^i,S_k^0)
	>
	\mathcal{N}(y_k^{i};\bar{y}_k^i,S_k^C)
	\label{14}
\end{align}

\begin{align}
	\frac{1}{\sqrt{2\pi S_k^{0}}}
	e^{-\frac{1}{2}\frac{(y_k^i-\bar{y}_k^{i})^{2}}{S_k^0}}
	&>
	\frac{1}{\sqrt{2\pi S_k^{C}}}
	e^{-\frac{1}{2}\frac{(y_k^i-\bar{y}_k^{i})^{2}}{S_k^C}}
	\label{15}
\end{align}

\begin{align}
	\implies
	\sqrt{\frac{S_k^{0}+\Sigma^{C}}{S_k^{0}}}
	&>
	e^{
		\frac{1}{2}(y_k^i-\bar{y}_k^{i})^{2}
		\left(
		\frac{1}{S_k^{0}}
		-
		\frac{1}{S_k^0+\Sigma^{C}}
		\right)}
	\label{16}
\end{align}

\begin{align}
	\ln(1+\phi_k)
	>
	(y_k^i-\bar{y}_k^{i})^{2}
	\left(
	\frac{\phi_k}{S_k^{0}(1+\phi_k)}
	\right)
	\label{17}
\end{align}

where
$\phi_k=\frac{\Sigma^{C}}{S_k^{0}}$
is the innovation ratio coefficient. It can be inferred that the
$i^{th}$ measurement dimension is outlier-free if

\begin{align}
	\frac{(y_k^i-\bar{y}_k^{i})^{2}}{S_k^{0}}
	<
	\ln(1+\phi_k)
	\left(
	1+\frac{1}{\phi_k}
	\right)
	\label{18}
\end{align}

The proposed method employs this test for identifying outliers in the
measurement vector and is therefore named the NIR-EKF. The estimator
singularly processes measurements to detect outliers, ensuring a
computational cost of $\mathcal{O}(m)$. Using a simple outlier detection
test instead of multiple hypotheses per measurements speeds up the
process, making NIR-EKF more computationally efficient than other
estimators, as shown in the next section.
\section{Simulation Results}

We compare our proposed estimators with leading outlier robust
methods: robust EKF (R-EKF) \cite{fang2021robust}, selective observation
rejection EKF (SOR-EKF) \cite{chughtai2022outlier}, and recursive outlier
robust EKF (ROR-EKF) \cite{piche2012recursive}, using the EKF as the
common inference tool for a fair comparison with MAP-based methods.
The reference methods are state of the art and have specifically been
developed for catering the problems of outliers in sensor measurements.
We consider a target tracking example with process dynamics as
\cite{chughtai2022outlier}

\begin{equation}\label{eqn19}
	\mathbf{x}_k =
	\begin{pmatrix}
		1&\Delta T&0&0\\
		0&1&0&0\\
		0&0&1&\Delta T\\
		0&0&0&1
	\end{pmatrix}
	\mathbf{x}_{k-1}+\mathbf{q}_{k-1}.
\end{equation}

The state vector $\mathbf{x}_k =
[a_k,\dot{a}_k,b_k,\dot{b}_k]^T$ has position coordinates
$(a_k,b_k)$ and the velocity coordinates
$(\dot{a}_k,\dot{b}_k)$. $\Delta T$ is the sampling period, which is set
equal to 1. Moreover, $\mathbf{q}_{k-1}\sim
\mathcal{N}(\mathbf{0},\mathbf{Q}_k)$ and
$\mathbf{Q}_k=0.1\mathbf{I}_{n\times n}$, where
$\mathbf{I}_{n\times n}$ is the identity matrix of size $n\times n$.
The selection of value of $\phi_k$ is crucial: a higher $\phi_k$
increases the detection threshold, possibly missing subtle outliers,
while a lower $\phi_k$ makes the method more sensitive, potentially
flagging normal variations as outliers. Consequently, we set
$\phi_k=10^6$.

Rewriting \eqref{eqn2},
\[
\mathbf{y}_k=\mathbf{s}_k+\mathbf{o}_k+\mathbf{r}_k,
\]
where
\[
\mathbf{s}_k =
\left[
s_k^{\theta_1},s_k^{\rho_1},
s_k^{\theta_2},s_k^{\rho_2},
\ldots,
s_k^{\theta_{\frac{m}{2}}},
s_k^{\rho_{\frac{m}{2}}}
\right]^T.
\]
The entries of $\mathbf{s}_k$, $s_k^{\theta_j}$, and
$s_k^{\rho_j}$ correspond to $j$th bearing and range sensors,
respectively, and are calculated as
$s_k^{\theta_j}=\atantwo(b_k-b^{\theta_j},a_k-a^{\theta_j})$ and
$s_k^{\rho_j}=
\left((a_k-a^{\rho_j})^2+(b_k-b^{\rho_j})^2\right)^\frac{1}{2}$,
where the angle readings are provided at coordinate
$(a^{\theta_j}=350(j-1),b^{\theta_j}=350(j\mathop{\rm mod}2))$
and range readings are obtained at
$(a^{\rho_i}=350(j-1),b^{\rho_i}=350((j-1)\mathop{\rm mod}2))$.

Also,
\[
\mathbf{o}_k =
\left[
o_k^{\theta_1}F(\lambda),
o_k^{\rho_1}F(\lambda),
\ldots,
o_k^{\theta_{\frac{m}{2}}}F(\lambda),
o_k^{\rho_{\frac{m}{2}}}F(\lambda)
\right]^T,
\]
where $F(\lambda)$ represents a Bernoulli distribution with parameter
$\lambda$ signifying the probability of an outlier present in the
reading. $F(\lambda)$ equals 1 when probability is equal to $\lambda$
and zero otherwise. Additionally,
$\mathbf{o}_k\sim\mathcal{N}(\mathbf{0}_m,\alpha\mathbf{R}_k)$,
where $\mathbf{0}_m$ denotes an $m\times1$ zero vector, and $\alpha$
is a scalar, signifying the strength of the outlier introduced.

Also,
$\mathbf{r}_k\sim\mathcal{N}(\mathbf{0}_m,\mathbf{R}_k)$, with
$\mathbf{R}_k$ being a diagonal matrix having entries
$r_\theta=10^{-3}$ and $r_\rho=5$.
Our system starts with an initial state defined as
$\mathbf{x}_0=[300,50,100,10]^T$.
We run the simulation for 100 Monte Carlo runs ($L=100$), 1000
time steps ($T=1000$), and $\alpha=800$. The comparison parameters
are time-average root mean-squared error (TRMSE) and elapsed time.
TRMSE is calculated by the following formula:

\begin{equation}\label{eqn20}
	\text{TRMSE}
	=
	\frac{1}{T}\sum_{k=1}^{T}
	\sqrt{
		\frac{1}{L}\sum_{l=1}^{L}
		(\mathbf{x}_k^l-\mathbf{\hat{x}}_k^l)^T
		(\mathbf{x}_k^l-\mathbf{\hat{x}}_k^l)
	}.
\end{equation}

where $\mathbf{x}_k^l$ represents the true value of the state vector at
time-step $k$ during the $l^{th}$ MC run and
$\mathbf{\hat{x}}_k^l$ represents the corresponding estimate from the
state estimation algorithm.

\begin{figure}[h]
	\includegraphics[width=0.9\linewidth]{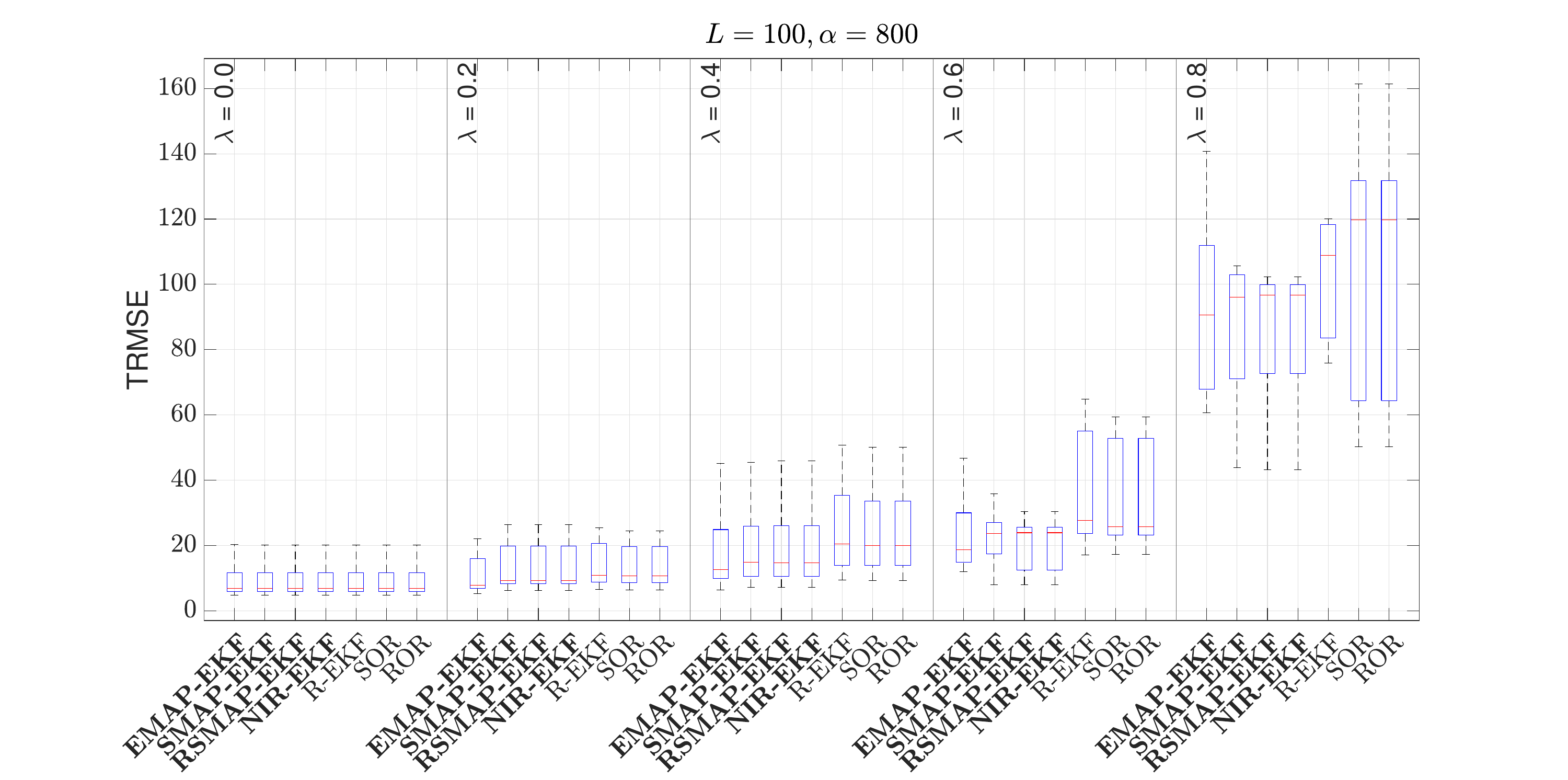}
	\caption{TRMSE versus different values of $\lambda$ for
		$m\in\{2,4,6,8,10\}$.}
	\label{TRMSE10E}
\end{figure}

\begin{figure}[h]
	\includegraphics[width=0.9\linewidth]{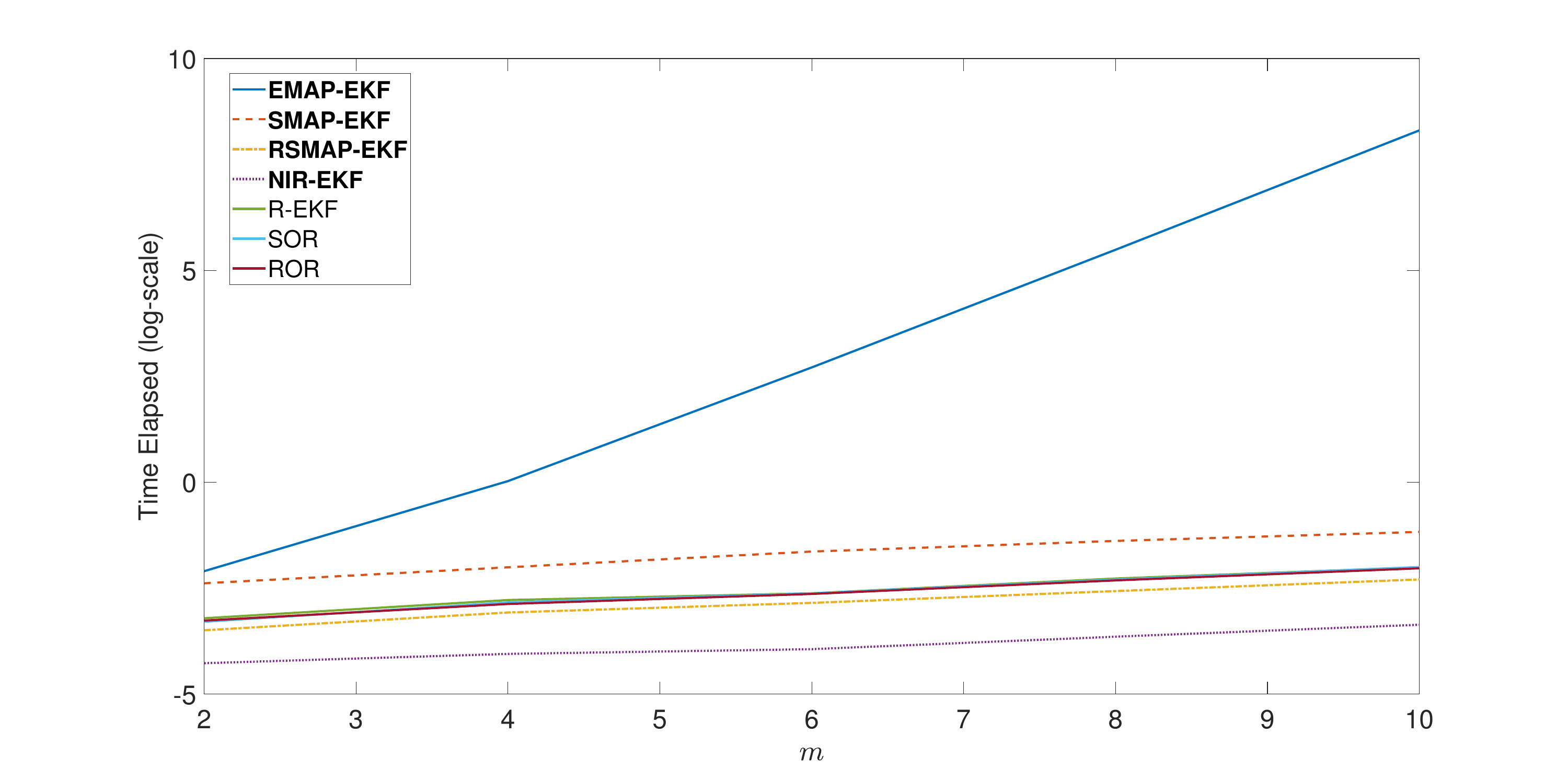}
	\caption{Log-scale computational time versus $m$.}
	\label{TEMAP}
\end{figure}

Initial simulations with
$m\in\{2,4,6,8,10\}$ shown in Fig.~\ref{TRMSE10E} boxplots that
EMAP-EKF has the lowest TRMSE values among all other methods, with
SMAP-EKF, RSMAP-EKF, and NIR-EKF performing similarly. We can also
observe that the performance of the MAP-based estimation techniques
is comparable to the other estimation methods for all values of
$\lambda$.

Fig.~\ref{TEMAP} shows the computational times for all algorithms
per run, emphasizing the EMAP-EKF's unsuitability for
computation-limited applications. A comparison of MAP-based methods
with other estimators reveals differences in running times.

\begin{figure}[h]
	\includegraphics[width=0.9\linewidth]{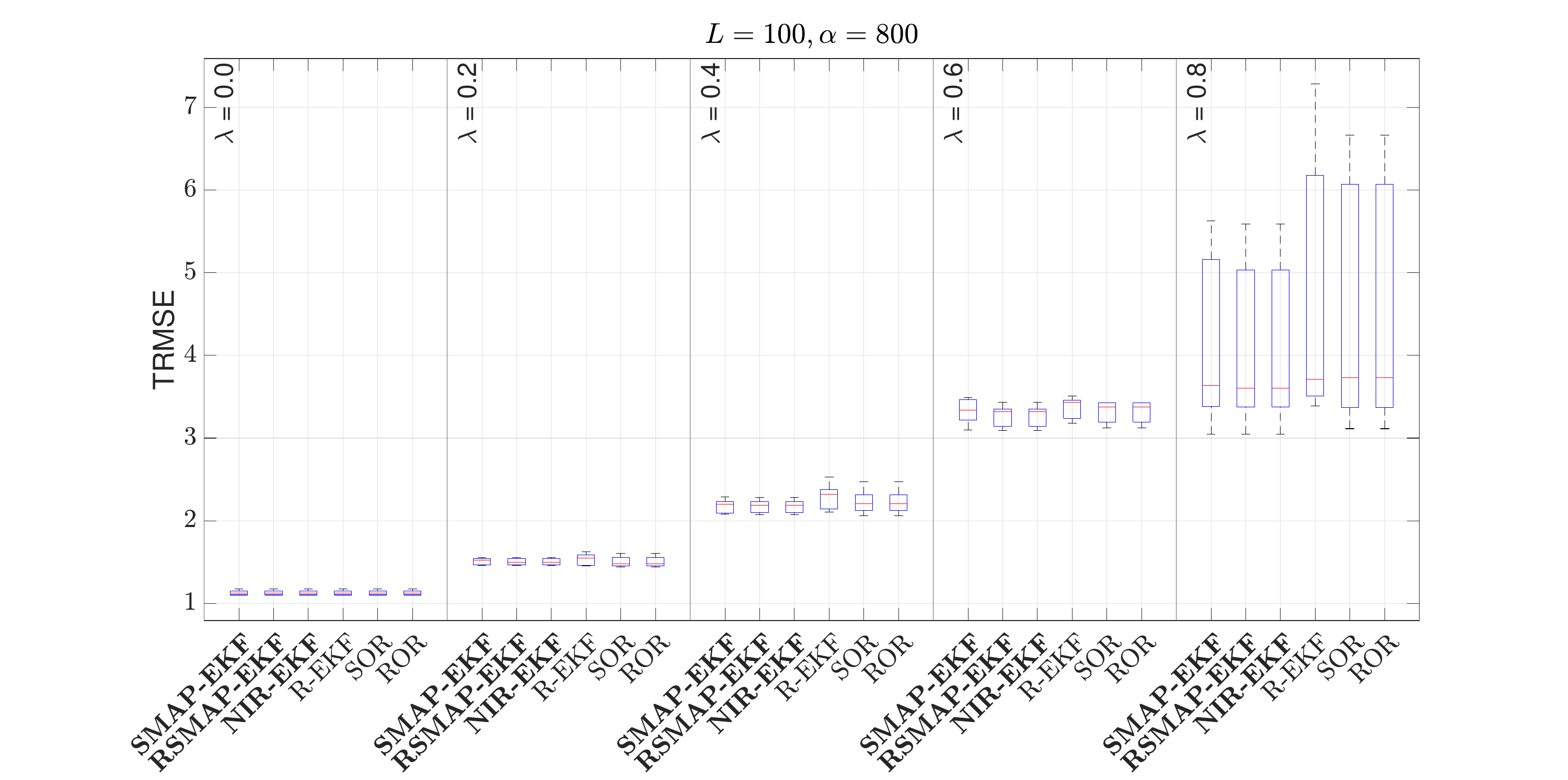}
	\caption{TRMSE versus different values of $\lambda$ for
		$m\in\{200,400,600,800,1000\}$.}
	\label{TRMSE1000E}
\end{figure}

\begin{figure}[h]
	\includegraphics[width=0.9\linewidth]{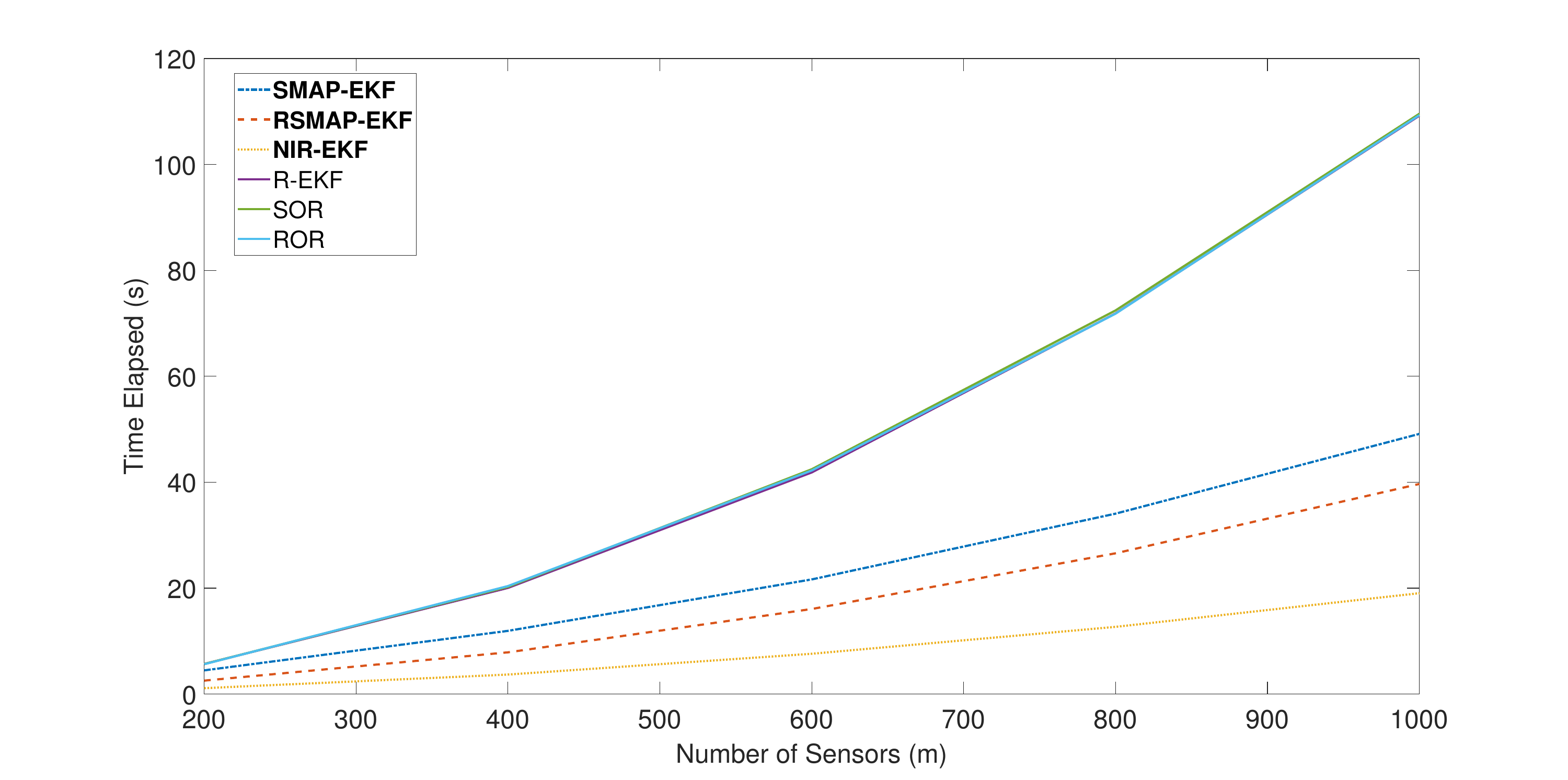}
	\caption{Elapsed time versus different values of $m$ for
		$\lambda\in\{0,0.2,0.4,0.6,0.8\}$.}
	\label{Time1000E}
\end{figure}

Further analysis from another simulation with
$m\in\{200,400,600,800,1000\}$ (shown in
Fig.~\ref{TRMSE1000E}) and results with
$\lambda\in\{0.0,0.2,0.4,0.6,0.8\}$, i.e., Fig.~\ref{Time1000E},
indicates that our methods match the performance of other robust
techniques but with less computational demand, as evidenced by the
lower times in Fig.~\ref{Time1000E}. This confirms that our MAP-based
methods are efficient yet effective alternatives to state-of-the-art,
robust techniques.

\section{Conclusion}

In this letter, we introduce MAP-based EKFs for robust state
estimation by efficiently handling outliers in sensor data occurring
simultaneously in multiple dimensions, leading to the development
of NIR-EKF. This
estimator demonstrates greater accuracy and enhanced computational
efficiency than other estimators and state-of-the-art, robust filters.
The simulation results highlight a significant improvement in the state
estimation quality, emphasizing the significance of our contributions
in managing multidimensional outliers.

\bibliographystyle{IEEEtran}
\bibliography{main}

\end{document}